# Essential System Services in Grids Dominated by Renewable Energy

*Market Reform Initiatives in Australia.*

Niraj Lal[1,2], Toby Price[1], Leon Kwek[1], Christopher Wilson[1], Farhad Billimoria[1,3], Trent Morrow[1], Matt Garbutt[4], and Dean Sharafi[1]

1 Australian Energy Market Operator, Australia
2 Australian National University, Centre for Sustainable Energy Systems, ACT, Australia
3 Oxford Institute for Energy Studies, Oxford University, Oxford, UK
4 Energy Security Board, Sydney, Australia

## A Case for Change

In 1863, a single arc lamp on Observatory Hill in Sydney was lit to celebrate the marriage of Prince Albert of Wales and Princess Alexandra of Denmark. It was the first use of electricity anywhere in Australia.

It took 25 years from this first light to the 1888 establishment of Australia's first permanent 240 V electrical grid in the small country town of Tamworth, New South Wales. Its two 18 kW DC coal-fired generators were supplied by the plentiful Gunnedah black coal basin nearby. In the same year, on the other side of the continent, C.J. Otte supplied electricity to the Western Australian Government House with a small 15 kW dynamo. By 1899, a full three-phase 240 V AC grid had been built on the east coast, establishing the foundation of the future power system across the country.

Then, as today, synchronous coal generators provided the majority of system services to maintain security and reliability. Those services included the inertia to maintain stable frequency, system strength to maintain stable voltage waveforms, and energy reserves to maintain the balance of supply and demand even with changing demand and unexpected contingency events.

Under this arrangement, the supply of these services has been conveniently tied to the supply of electrical energy with synchronous generators providing support simply by being synchronized with the electric grid. For over a century, as the electricity infrastructure and trading systems grew, no separate mechanisms were developed to manage these "ancillary services" to the power system. Instead, grid connection standards implicitly regulated an equitable division of costs among facilities in rough proportion to their size. Operators could recover these "costs of doing business" as part of their energy revenue.

The generation mix around the world is changing rapidly. In Australia, this is happening at a world-leading rate, from having the third most carbon-intensive electricity generation in the world in 2010 to now regularly receiving more than a third of its electricity from renewables. One in five households in the country have distributed photovoltaics (PV) systems (at an average of 600 W installed per person, growing at 250 W per person per year)–the highest rate of PV uptake in the world. At times, over 100% maximum instantaneous solar and wind penetration is achieved in some regions.

Solar and wind generators connect to the AC grid via power-electronics-based inverters, which do not provide traditional system services by default. This means that while inverter-based resources can replace the *energy* previously provided by synchronous coal and gas generation, the provision of *system services* is not replaced in proportion.

The remaining fleet of synchronous resources then faces a growing burden of providing system support services, such as frequency and voltage control and spinning reserves, while revenues simultaneously fall with electricity prices and reduced market share or energy generation. Left unchecked, this dynamic undermines the implicit stability that has historically supported the electricity system.

In Australia's National Electricity Market (NEM), this has manifested in a 10-fold increase over the past five years in the number of occasions the system operator has had to intervene outside of normal market operations to maintain security and reliability (Figure 1a). There has been a significant reduction in frequency control performance since 2007 (Figure 1b) due to reduced provision of primary frequency control. Uncertainty and variability in net demand from increasing renewable penetration are expected to triple in the NEM over the coming five years (Figure 1c) as solar and wind are projected to regularly meet 100% of demand (Figure 1d).

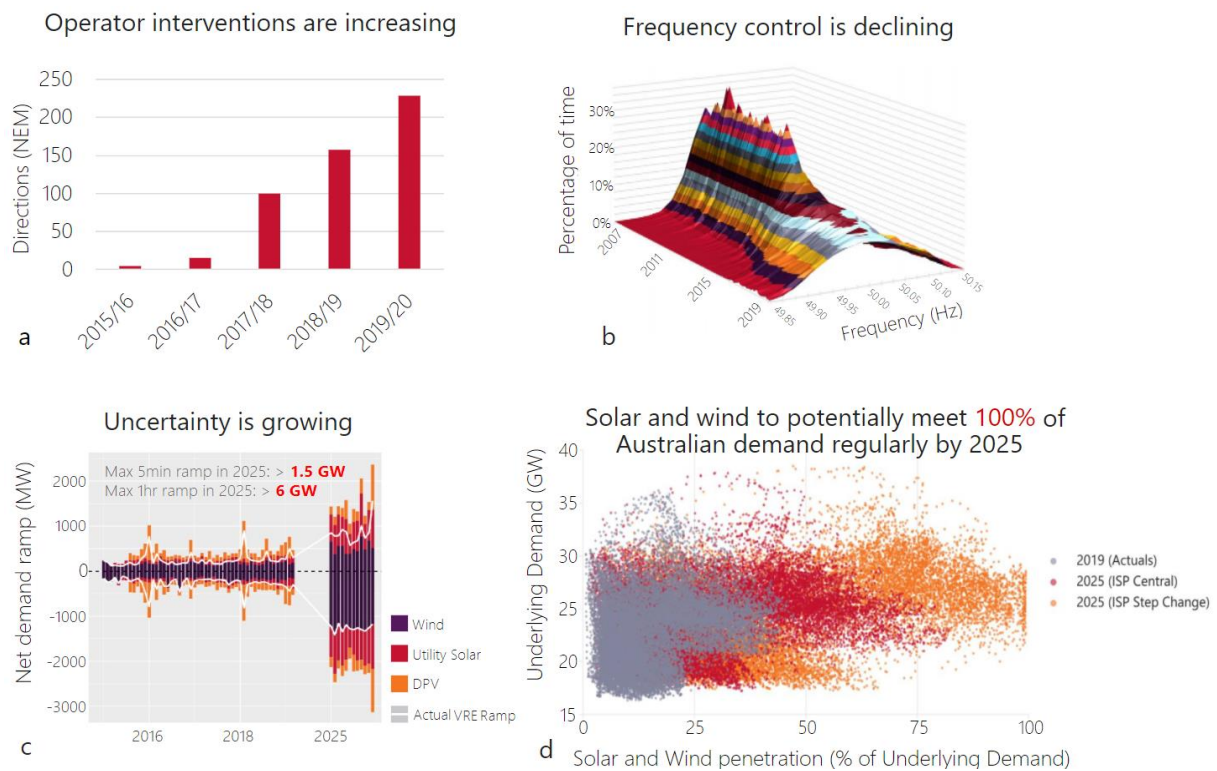

*Figure 1 a) Operator directions in the NEM; b) frequency distribution plot in the NEM to 2019; c) butterfly plot of 5-minute net demand ramps (historical and forecast); d) forecast penetration of solar and wind as a percentage of underlying demand. (Source: AEMO Renewable Integration Study Stage 1 2020, AEMO Frequency and Time monitoring report Q1 2020)*

As the generation mix has changed, a handful of events have catalyzed political interest and action. After a September 2016 state-wide blackout in South Australia, the Australian government commissioned a "Review of the Future Security of the National Electricity Market" by Australia's chief scientist Alan Finkel. This led to the establishment of an overarching Energy Security Board (ESB) to implement a "long-term, fit-for-purpose market framework" to deliver a "secure, reliable and lower emissions electricity system at least cost" in the NEM.

A key workstream of this reform program is to establish new markets and mechanisms for providing system support services. These were traditionally called 'ancillary' services but are now increasingly being referred to as essential system services (ESS) in recognition of their changing value in grids with low levels of synchronous generation.

There is a growing consensus that without market reform, the market operator's remit to "keep the lights on" will likely be accompanied by increased curtailment of renewables and increased complexity of operation. This trend is already being observed. In 2019-20, renewables were curtailed on average 7% of the year in the NEM due to ancillary service requirements, and operator interventions were in place more than 10% of the year.

This article presents the Australian approach to the challenge of providing ESS in grids with a very high penetration of renewables, outlining first the physical and regulatory contexts of two comparative systems and markets – the NEM (with five regions across Australia's eastern and southern states), and Western Australia's wholesale electricity market (WEM) – and their concurrent programs of reform.

The changing nature of ESS in both markets, principles of market design, the spectrum of opportunity for procurement of various new services, and the integration and congruency challenges of addressing them holistically are then discussed. Finally, we present the Australian pathway of reform and a vision for the future of ESS with the hope that it may prove helpful for other countries across the world on similar decarbonization pathways.

## Context

Australia's electricity networks span vast distances across a continent roughly the area of the United States, but with less than a 10th of the population.

The energy industry, historically government-owned, was deregulated in the 1990s to disaggregate the vertically integrated state utilities and support competition. This allowed cross-border trading of electricity between states and territories.

The NEM was established in 1998. The isolated nature of the Western Australia and Northern Territory electricity systems was a significant barrier to continent-wide integration of both infrastructure and policy. It was only in 2006 that Western Australia's WEM was established, covering the southwest region of the state and serviced by the South West Interconnected System, spanning an area roughly the size of the United Kingdom. The interconnected NEM power system is serviced by approximately 40,000 km of transmission network. The islanded South West system integrates approximately 7,800 km of transmission network.

## The Australian Energy Markets

Arising from a period of widescale deregulation, the NEM was established with a strong commitment to market efficiency with real-time five-minute dispatch intervals, no day-ahead capacity markets, and very high market-price caps (currently $1,000/MWh to $15,000/MWh). In 2021, the settlement time will reduce from 30 minutes to five minutes to align with dispatch, further sharpening market efficiency in the continuous matching of electricity supply and demand. Along with providing efficient incentives for participants, real-time price mechanisms also allow the possibility of contracts for difference and hedging to support long-term contracts and risk mitigation. This is extensively conducted in the NEM through hedging and swap contracts.

NEM markets for system services were set up with a similar commitment to real-time pricing (for those services that were remunerated). The NEM has six frequency response markets (frequency control ancillary services): contingency frequency response raise and lower services for each of six seconds, 60 seconds, and five minutes response times, and an additional causer-pays primary frequency response service. There are additional non-market services for network support and control (such as transient oscillation control) and system restart.

Reflective of its smaller and more concentrated nature, the WEM balances market efficiency with greater structured procurement, including a capacity market (the Reserve Capacity Mechanism), a day-ahead energy market, the short-term energy market, and a real-time energy market with 30-minute dispatch (with lower market-price caps, currently $1,000/MWh to $382/MWh).

For system services, the WEM prioritizes structured procurement via a regulation market (load-following ancillary services) and other system services procured under contract, including frequency response (spinning reserve) and, like the NEM, similar non-market services for network control and system restart.

## The Post-2025 Program

In 2019, the Australian federal and state and territory governments asked the ESB to advise on a long-term, fit-for-purpose market design for the NEM that could apply from 2025 in response to the profound energy transformation occurring across the country.

The initiative has become known as the "Post-2025 Market Design Project," focusing on the entire energy supply chain – from the wholesale energy market through transmission and distribution to behind-the-meter distributed energy resources. The ESB, resourced collaboratively by the Australian Energy Market Commission, Australian Energy Market Operator (AEMO) and Australian Energy Regulator working together with ESB staff, set up four workstreams to consider the issues and develop potential solutions:

- Resource adequacy through the transition
- Essential system services and scheduling and ahead mechanisms
- Demand-side participation
- Access and transmission

Industry and customer stakeholders have been extensively involved and consulted, and there is broad recognition that the individual workstreams are intrinsically interrelated and must be considered together for a coherent whole design.

There is a wide range of views about each of the workstreams, but responses indicate the reform of system service provision is the highest priority and most urgent. Such reform needs to occur before 2025 to address tighter frequency control, structured procurement for synchronous generation commitment (for system strength and inertia) potentially combined with an ahead mechanism to support scheduling, and the exploration of possible operating reserve and inertia spot markets.

## WA 2022 Program

In 2019, the Western Australia government formed the energy transformation task force charged with making clear policy decisions with robust consultation to ensure coherent reform for a full overhaul of the market regulatory framework to go live in 2022. The task force has an explicit focus on the assessment and redevelopment of a new ESS framework.

# Box: Essential System Services

All power systems require a suite of system services, traditionally known as ancillary services but increasingly referred to as "essential system services (ESS)," necessary for the secure and reliable operation of the system. Services can often perform the same function but vary in their names, implementation, competitiveness, and remuneration mechanisms across jurisdictions. Below is a summary of the various services that exist in Australia with their WEM and NEM implementations.

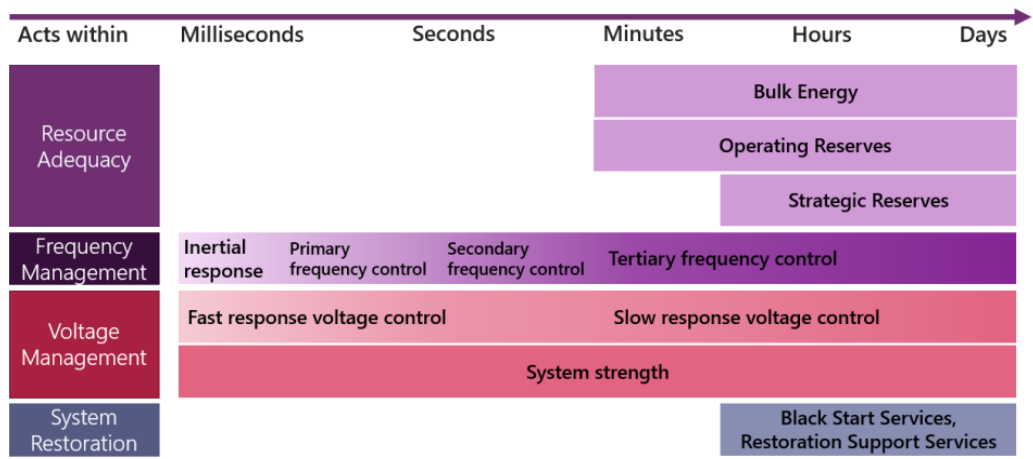

*Figure: Operation timescales and categorization of certain ESS, Source: AEMO Power System Requirements, 2020*

| Service | Description | NEM Equivalent | WEM Equivalent |
| --- | --- | --- | --- |
| Bulk Energy | Energy to meet demand (both scheduled and unscheduled) | Energy (5m dispatch, 5m settlement [from 2021]) | Energy (30-minute dispatch and settlement). Moving to 5-minute dispatch 2022 and 5-minute settlement in 2025 |
| Regulation | Maintains frequency within the normal operating band, operating within seconds | Regulation Raise/Lower | Load-Following Ancillary Service (LFAS) Up/Down market<br>Moving to co-optimized Regulation Service, 2022 |
| Primary Frequency Response (PFR) | Arrests and stabilizes frequency after following an event which results in a sudden mismatch of demand and supply, operating within milliseconds | Droop response and Fast Raise/Lower (6s).<br><br>Possible new Fast Frequency Response (<2s) from 2022. | Droop Response and Spinning Reserve<br><br>Moving to co-optimized Contingency Reserve real-time market, 2022 |
| Secondary Frequency Response (SFR) | Restores frequency to its normal operating band after an event operating within seconds to minutes | Slow Raise/Lower (60s) and Delayed Raise/Lower (5m).<br><br>Possible combination of 6s and 60s services from 2022. | Spinning Reserve<br><br>Moving to co-optimized Contingency Reserve real-time market, 2022 |
| Tertiary Frequency Response (TFR) | Reschedules / unloads facilities that provide primary and secondary frequency response so that they available to again respond to a new event | Energy redispatch | Energy redispatch, and redispatch of the government-owned energy assets[1]Moving to co-optimized Contingency Reserve real-time market, 2022 |
| Inertia Service | Physical inertia that reduces the rate of change of frequency (ROCOF) following a contingency event | No existing service<br><br>Possible scheduling of synchronous resources through a Unit Commitment for Security mechanism or Synchronous Services Market.<br><br>Possible future Inertia spot market. | No existing service<br><br>Moving to a co-optimized ROCOF Control Service, 2022 |
| Operating Reserve | Balances the supply and demand of energy over a minute to hours horizon | Possible new market for Operating Reserves or Ramping Availability from 2025 | No explicit service. Managed by energy redispatch and self-commitment |
| System Restart | Facility capability to restart a black system, and to assist in its reconstruction following a black system event | System Restart Ancillary Service (SRAS) | System Restart Service (SRS)<br><br>Provided as part of Non-Co-Optimised Essential Systems Services Framework, 2022 |
| Voltage Support and System Strength (discussed further in text) | Stabilizes voltage in a location of a network | Network Support and Control Ancillary Service (NSCAS)<br><br>Possible scheduling of synchronous resources through a Unit Commitment for Security Mechanism or Synchronous Services Market. | Network Control Service (NCS)<br><br>Provided as part of Non-Co-Optimised Essential Systems Services Framework, 2022 |
| Capacity | Procurement of capacity (both generation and demand-side management) to meet forecast peak demand on the yearly time horizon | No explicit service except for Reliability and Emergency Reserve Trader function.<br><br>Possible new market for Operating Reserves or Ramping Availability in the NEM | Reserve Capacity Mechanism. Annually administered price mechanism for certified capacity. |

*Table 1: A summary of various ESS in Australia and their implementations within the WEM and the NEM.*

## Principles for Procurement and Market Design

For the impending challenge of redesigning procurement frameworks for ESS, it helps to first consider broad principles of market design alongside the intrinsic valuation of power system security. The objective of procurement frameworks should be to create efficient and effective economic mechanisms to deliver operational requirements.. The operational requirements of power system security must focus on the management of the underlying physics of an electrical network with sufficient redundancy and robustness in the face of uncertainty and risk.

## Market Design for ESS

A recent report compiled by FTI Consulting for the ESB highlighted seven principles important to the design of effective procurement frameworks for ESS (see Figure 2 Principles of market design for essential system services. Source: Adapted from FTI Consulting Report to the ESB, 2020.).

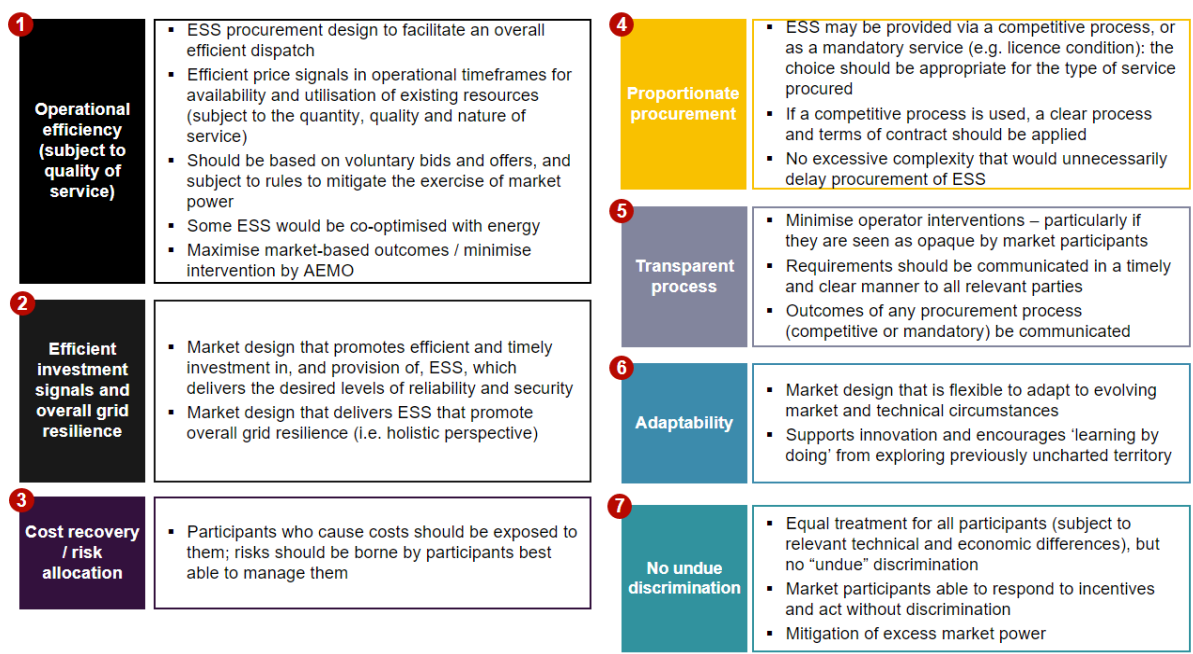

Figure 2 Principles of market design for essential system services. Source: Adapted from FTI Consulting Report to the ESB, 2020.

These principles are fundamental in framing the design problem from a regulatory and market perspective. Alongside these principles is recognition that any design process necessarily involves a compromise between elements to achieve an overall workable design.

In particular, there is a natural tension between the idealized theoretical design of markets with assumptions of economically rational behaviour and the physical reality of operation which is complex, uncertain, non-linear and failure prone. There are additional asymmetric costs of market efficiency and market failure. While designers may prefer complex, multi-layered, and co-optimized markets, operators may prefer conservative, expensive, and un-optimized solutions. Striking the right balance to develop efficient and robust economic solutions to technical challenges requires the rigorous and combined efforts of power system engineers and economists.

Policy makers have a variety of regulatory and market instruments available to them. Options include technical standards or licenses, operational directions or interventions, regulatory delegations (such as to network monopolies or other central agencies), individual contracts with providers, ESS auctions, and tenders and short-term spot markets.

Regulated approaches can provide greater comfort in the technical provision, especially given technically complex security services (such as system strength). While market approaches provide the opportunity for greater efficiency, there is potential for financial innovation to out-compete technological innovation. Market solutions can also optimize against the technical specification of a service, creating a lack of resilience.

A case in point is the design of contingency frequency response markets in the NEM, where technical specifications guided by normal operating frequency bounds enabled wide frequency governor dead bands. In the face of uncertainty, this led to poor frequency performance and system fragility, only recently corrected by re-implementation of stringent mandatory primary frequency response requirements. By contrast, the WEM complements a spot market for regulation services with an obligatory droop requirement, which has led to improvements in frequency management.

Trade-offs abound for investment considerations given commercial risk appetite. While spot markets, if appropriately designed, can provide efficient scarcity price signals, investment decisions on long-duration assets are typically made in the context of longer-term revenue and cash flow visibility.

In the design of ESS, it is relevant to consider:

- Framework flexibility is needed in managing current principles of provision (such as from synchronous generators and synchronous condensers) while accommodating future innovation (inverters providing "synthetic inertia" and grid-forming capability).
- The locational nature of service-provision. For example, fault current and system strength are highly locational relative to inertial frequency response which is system-wide.
- The complexity of co-optimization in the context of uncertainty.
- The challenge of valuing ESS and the consequent difficulty of allowing procurement quantities beyond minimum levels to provide additional robustness or resilience.
- The trade-offs of operational complexity and market sophistication: Complex markets create more points of failure.

During this period of rapid change, adaptive governance and procurement approaches are helpful. For ESS, a flexible contractual framework would support the operator to mitigate fast-evolving system risks, potentially accompanied with an adaptive regulatory change process that supports participant decision-making.

## Other International Approaches

While Australia's power system finds itself in uncharted territory in the penetration of variable renewable energy (VRE) and distributed solar, there are pioneering advances in market design for system services being explored concurrently across the world. This section reviews some key developments in comparable systems in the United Kingdom and United States.

In the United Kingdom, electricity systems operation and procurement of system services are delegated to National Grid Electricity System Operator (NGESO), a subsidiary of the for-profit private National Grid UK, which also owns and operates the transmission network. This framework provides a comparatively high degree of flexibility in the approach to procurement with NGESO utilizing competitive tenders of varying duration and structure in procuring services.

Standardized system-wide frequency and reserve products have contributed to shorter-term frequent contract auctions, while more individually tailored and longer-term contracts were used to secure requisite investment for services with locational requirements and smaller provider pools. A recent initiative is the Stability Pathfinder tender, which procured a combination of services including fault-levels and inertia. Reactive power, traditionally an obligatory service, is also increasingly procured through competitive tender approaches.

NGESO is subject to a unique financial incentive scheme with payments based on performance evaluated by the regulator Office of Gas and Electricity Markets (Ofgem) through an annual score card assessment. The discretion provided to Ofgem has been particularly useful in a rapid-change environment, providing flexibility to respond to evolving technical scarcities and to modify and adapt procurement on an ongoing basis. This has also left NGESO to deal with the issue of supporting investment by initially procuring newer services via longer-term contracts (to underpin investment), moving towards shorter-term auctions as business models become established.

By contrast, regulatory regimes in the United States and Ontario, Canada, have delegated system services to independent system operators (ISOs), which are not-for-profit entities with relatively less discretion to make decisions on ESS procurement. Procurement approaches tend to be codified in regulations with changes subject to detailed review, stakeholder engagement, market participant vote, and, in some cases, approval of the Federal Energy Regulatory Commission.

Given the need for transparency, ESS have tended to be procured via either short-term spot markets (predominantly frequency and reserve products) or mandatory provision. Spot markets have provided transparency and visibility of pricing, allowing financial markets to develop around services underpinning investment. However, given regulatory structures, incentive mechanisms for U.S. ISOs have proven to be challenging due to narrow incentive thresholds and forecasted delivery targets. In practice, these challenges, combined with the regulatory processes, have limited the ability of ISOs to develop new products expediently.

While many jurisdictions are adapting technical standards for inverters, there has been less emphasis to date in international jurisdictions on service procurement concerning system strength.

The meshed nature of North American grids means system strength and the provision of fault current is of less concern from a technical perspective, and as a result, it is not explicitly defined as a system service for many regions, including the New York ISO, Mid-Continent ISO, and Ontario Independent Electricity System Operator.

Australian market designs have strong parallels with security-constrained gross power pool models common across markets in North America, apart from procedures for centralized unit commitment and two-settlement market clearings which are not part of the NEM. However, given the extent of VRE penetration and the unique operational phenomena observed in Australian grids, the grids will likely have to forge novel approaches to procure these complex and multi-faceted technical services. These approaches will also have to work alongside the broader challenge of a 5-minute spot market framework without ahead or capacity markets.

# Spectrum of Opportunity

## Procurement Frameworks

Having identified a case for change and reviewed the principles of market design for ESS, the challenge then progresses to canvassing the "spectrum of opportunity" in resolving the missing services that arise as inverter-based resources (IBR) replace synchronous generators.

There are many options to procure ESS, but frameworks can be broadly categorized along an axis of market efficiency (see Figure 3 A spectrum of opportunity for ESS in the NEM and WEM, indicating the status of current ESS market mechanisms, with an implicit axis of "greater market efficiency" towards the right. Source: Adapted from FTI Consulting Report to the ESB, 2020.):

1. "Market operator interventions and self-provision of services" without market-based remuneration (currently used for system strength, inertia, and operating reserves).

2. "Structured procurement" via non-spot market mechanisms (currently used for emergency out-of-market reserves, voltage control, and network support/control).

3. "Spot market-based" provision of services (currently employed for energy, regulation, and contingency frequency control).

Although there is a preference for real-time signaling, not all system services are suited for market-based procurement. The assessment of market design for each service includes factors such as the measurability/fungibility of the product, competition and co-optimization scope, complexity and simplicity, and locationality.

This section introduces various options for market design for each stream of ESS under consideration, namely operating reserves, frequency management, synchronous services, and inertia.

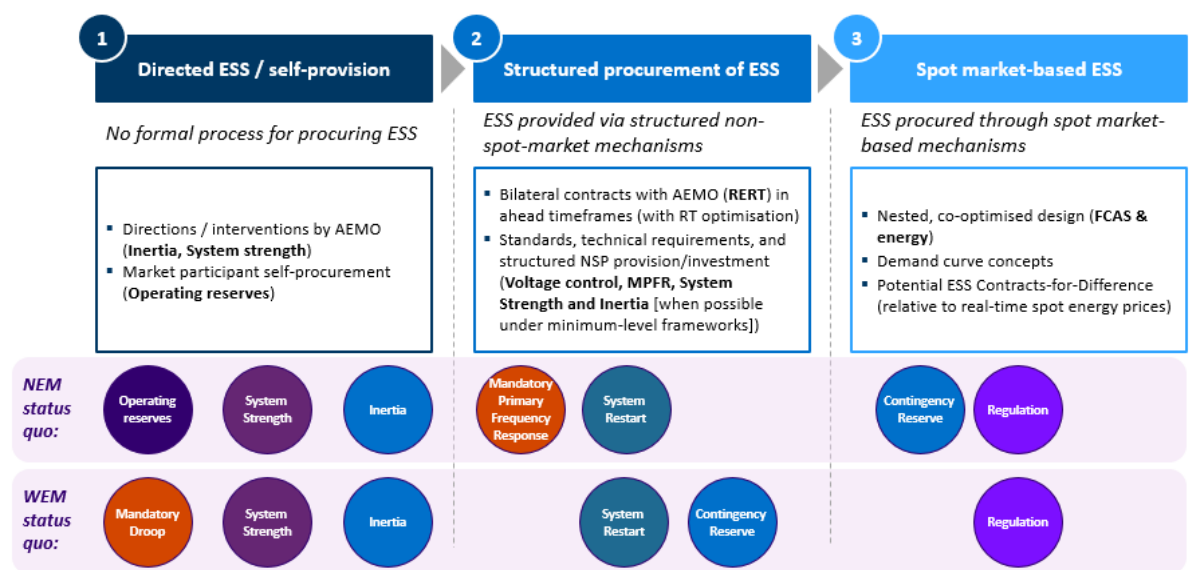

*Figure 3 A spectrum of opportunity for ESS in the NEM and WEM, indicating the status of current ESS market mechanisms, with an implicit axis of "greater market efficiency" towards the right. Source: Adapted from FTI Consulting Report to the ESB, 2020.*

## Operating Reserves

Energy markets must maintain supply and demand in instantaneous balance with prices set through a spot market. Market participants often have separate contracts across their portfolio to manage the risk around the energy spot price. The market operator, however, typically does not see these contracts. Instead, it must rely on faith in participants displaying economically rational behaviour and taking advantage of high prices at times of supply scarcity.

This faith is increasingly being tested by the changing nature of generation, with the "invisibility" of behind-the-meter distributed PV generation, and the variability and uncertainty of large-scale wind and solar (Figure 4 a) The probability distribution of 30-minute forecast errors for South Australia, Summer 2019-20, 2-6 pm.; b) probability curve that the 30-minute forecast error is higher than any particular level of reserve, which may inform an efficient reserve demand curve; c) an example 30-minute ramping "availability" product to address unexpected ramps over a 30-minute time horizon, adapted from Brattle Consulting Report to AEMO, 2020.a). The likely result of this uncertainty is the system operator managing the system more conservatively, leading to greater VRE curtailment as risk becomes excessively high. The possible design of an operating reserve or ramping availability service under current consideration may help address this challenge in the NEM.

There are several market options to procure operating reserves including a) procuring firm "availability" in the dispatch interval 30 minutes ahead (Figure 4 a) The probability distribution of 30-minute forecast errors for South Australia, Summer 2019-20, 2-6 pm.; b) probability curve that the 30-minute forecast error is higher than any particular level of reserve, which may inform an efficient reserve demand curve; c) an example 30-minute ramping "availability" product to address unexpected ramps over a 30-minute time horizon, adapted from Brattle Consulting Report to AEMO, 2020.c); b) holding a certain level of spinning "callable" reserve available to be triggered to dispatch as energy; or c) procuring operating reserve "headroom" in the coming 5 minutes to dispatch as energy. With each option, the use of a demand curve constructed from historical forecast errors may inform the efficient level of reserve or firm availability to procure (Figure 4 a) The probability distribution of 30-minute forecast errors for South Australia, Summer 2019-20, 2-6 pm.; b) probability curve that the 30-minute forecast error is higher than any particular level of reserve, which may inform an efficient reserve demand curve; c) an example 30-minute ramping "availability" product to address unexpected ramps over a 30-minute time horizon, adapted from Brattle Consulting Report to AEMO, 2020.b).

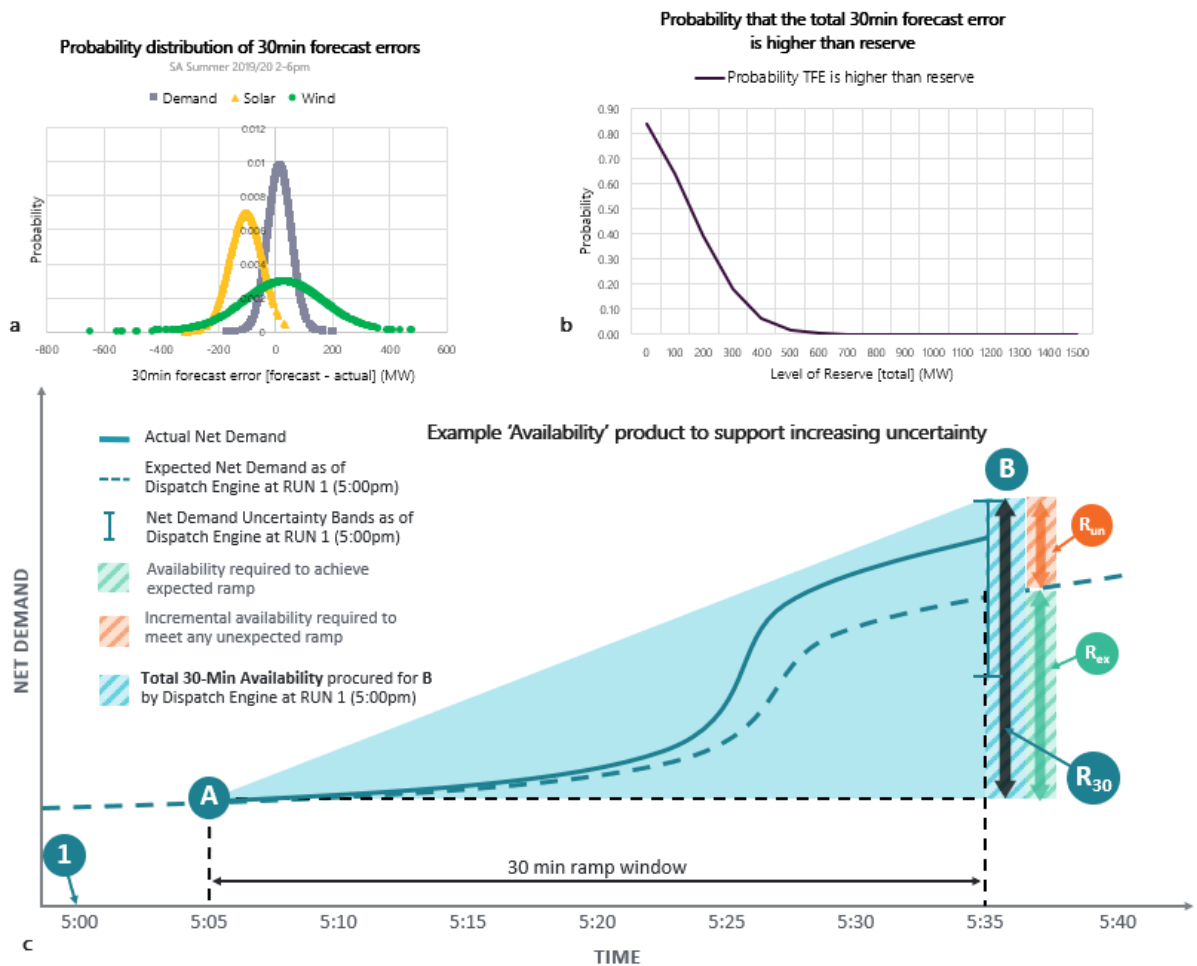

*Figure 4 a) The probability distribution of 30-minute forecast errors for South Australia, Summer 2019-20, 2-6 pm.; b) probability curve that the 30-minute forecast error is higher than any particular level of reserve, which may inform an efficient reserve demand curve; c) an example 30-minute ramping "availability" product to address unexpected ramps over a 30-minute time horizon, adapted from Brattle Consulting Report to AEMO, 2020.*

These options are being developed for possible NEM implementation in the next two years. Decisions on a final preferable new market will be based on trade-offs between operator confidence, market efficiencies, and potential adverse impacts on the energy spot market.

### Frequency Management

This class of services encompasses the need to schedule reserves of energy capacity that respond to unexpected changes in the load-generation balance (in addition to synchronous inertia discussed below). There are two broad categories to consider:

- Regulation reserves: Responding to ongoing and smaller imbalances, primarily due to variations in demand, and generation from intermittent sources.
- Contingency reserves: Responding to sudden and very large disturbances, such as the loss of a major generation unit.

Under assumptions of reasonable connectivity and system strength, frequency management can be sourced from any network location. Much like in the case of standard energy supply, this "global pool" of resources lends itself to procurement via a centralized, co-optimized spot market (energy dispatch can be considered a very slow class of frequency control). However, the desire for a universal and highly optimized market design must be carefully weighed against the complexities and irregularities this can create in a physical system.

Figure 5 illustrates this consideration through a system operations abstraction of frequency management for contingency response services. In this view, the physical response of the entire generation fleet is aggregated and considered according to different performance requirements for deployment and sustainability of power output into inertial, primary, secondary, and tertiary response. These distinctions are not fundamental, but reflect control structures formed around physical properties and useful trade-offs optimized in the allocation of power system resources.

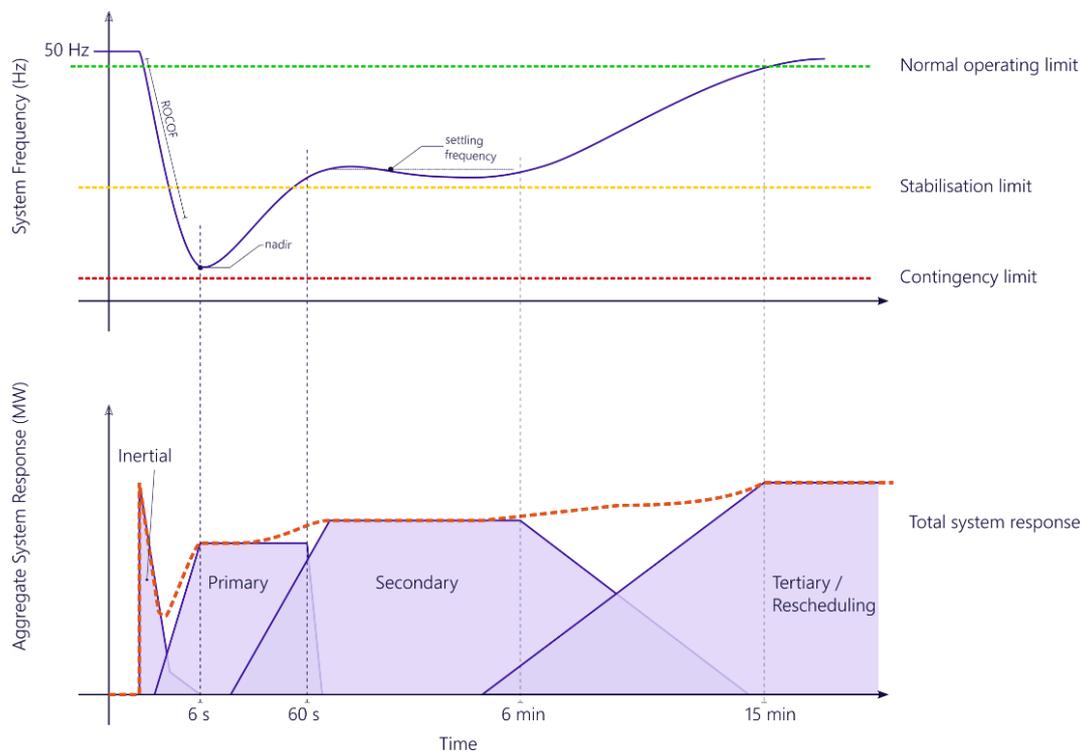

*Figure 5 Top: System frequency appropriately managed after a contingency event; Bottom: Frequency management mechanisms to support the restoration of system frequency following a contingency event.*

Three critical security limits must be managed following a generation contingency (Table 2).

*Table 2: Frequency limits to be managed following a generation contingency.*

| Limit | Description | Management |
|---|---|---|
| ROCOF | Maximum rate of change of frequency (ROCOF) in the first 1-2s | Synchronous inertia, and potentially "virtual inertia" from power electronics resources |
| Nadir | Absolute minimum frequency, typically reached around 6s | Primary response of local generation control systems |
| Settling frequency | A "quasi-steady-state" frequency maintained while the system is restored to normal operating conditions | Secondary response directed by central generation control schemes |

Exact operating limits vary due to jurisdictional norms and reliability standards, however, these standards ultimately reflect the physical tolerances of an electrical plant. Inverter-based facilities, for example, generally have a higher tolerance to ROCOF than rotating machinery. The ideal procurement model also incorporates incentives to reward tolerances. It reduces overall service requirements in addition to procurement of suppliers. The NEM reform program is reviewing the feasibility of including additional fast frequency contingency response (with response times of less than 1 second) alongside mechanisms to support the efficient provision of (currently mandated) primary frequency response within the normal operating frequency band.

### Inertia and System Strength

The procurement of synchronous services, namely system strength and inertia, is particularly complex to transition away from their traditional provision as a by-product of generation from synchronous generators. Options to replace this include building additional synchronous resources by the network operator (for example, synchronous condensers with flywheels), or creating incentives for the provision of services through advanced power electronics [See Box: Australia's Big Battery].

There is an opportunity to procure inertia as a separate service, an option being implemented as part of the market reform in Western Australia through the ROCOF Control Service [See Box: WA ROCOF Control Service].

# Box: WA ROCOF Control Service

In August 2019, the Western Australian energy transformation task force found that a real-time co-optimization of all frequency control services–including inertia–was most appropriate for the future WEM, driven by a mixture of physical, operational, and market considerations.

Historically, the WEM relied on an empirically derived rule of thumb: 70% of the largest generation contingency (in MW) was allocated as headroom across a set of designated facilities. Analysis and comparison of this approach identified that the combination of isolation and relatively small size resulted in the WEM being run close to its technical limits.

The transition to a greater penetration of renewables has necessitated a more sophisticated market design and led to a preference for a real-time spot market to optimize both system inertia and primary response speed. In this context, initial design options focused on the correct balance of service definition "segmentation," for example, adding a 1-second, 2-second, and 3-second service to complement the co-optimized 6-second and 60-second markets as done in the NEM. With system requirements abstracted to fundamental quantities (i.e., generic MW specifications), the optimal delivery of these services would be by market dynamics, irrespective of the underlying technology.

Unfortunately, investigations and analysis revealed issues with the multi-segment approach from both the physical and market perspectives:

Physical:

- Each segment adds complexity and increases the degree of "fantasy" space in which the commercial abstraction diverges from physical reality. In practice, there is no clean, linear separation of MW into convenient buckets.
- Further, inertia is only superficially the same as primary response. True rotating machinery has a fundamentally instantaneous response while power electronics suffer from a delay in electronic detection on the same order (<1 s) of the critical ROCOF period.

Market:

- Each segment adds complexity, resulting in additional infrastructure/systems overhead plus opportunity to game/manipulate market systems.
- Especially in a relatively "shallow" market (pool of suppliers), more complexity increases the chances of a participant effectively exercising power over a market.

The task force decided a single segment was most appropriate. While multiple segments allow for more service differentiation, in practice these gains were marginal while the downsides were guaranteed.

Implementation of this direction required a fundamental change in the perspective of service definitions. Rather than split physical responses across multiple segments, the entire response profile is characterized in reference to a perfect exponential response (see figure A) chosen to approximate the output of a physical turbine. The response factor is then converted into a multiplier that incentivises speed.

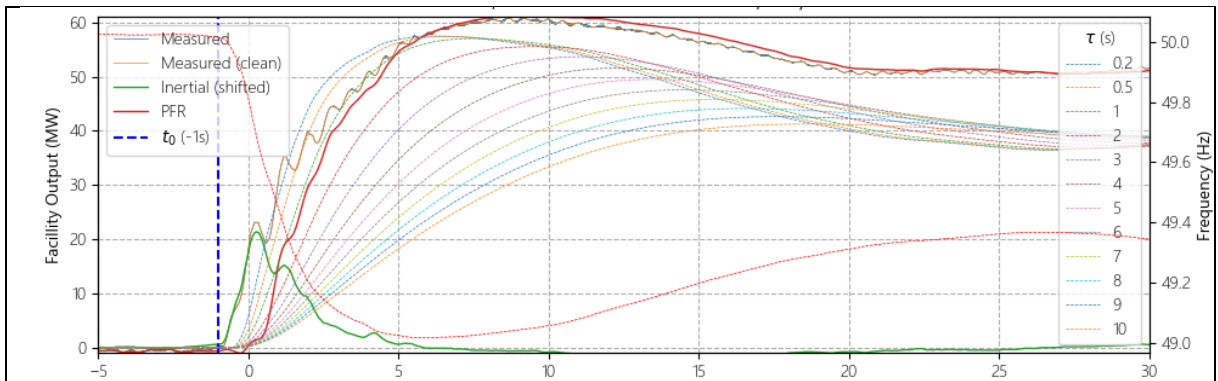

Figure A: The physical response of a gas turbine is measured and compared against an array of hypothetical "perfect exponential" responses of different speed.

Inertia is split from the primary response in recognition of the underlying physical differences, while inverter-based generation is credited through very high performance-multipliers.

The task force, however, noted the ongoing research and development of inverter-based technology, and named the inertial service "ROCOF Control" in recognition that future developments may open this segment to power electronic devices.

System strength is an emerging concept broadly defined as the strength of a power system's voltage waveform. It is closely associated with both inertia and fault-current levels but not comprised solely of either. The ability to maintain a stable waveform is decreasing as inverter-based resources connect to the system. The appropriate procurement mix for system strength may incorporate elements of various frameworks with challenges for policy and regulation in appropriately allocating risks, costs, and benefits to customers, the system operator, and network service providers.

A possible approach to procurement is to mandate threshold levels at all nodes across the network (via the specification of a minimum fault current level or short-circuit current ratios) and allocate maintaining these levels to the transmission network service provider. As regulated entities, there is some incentive for providers in procuring capital equipment to include on their regulated asset base. This may discourage the provision of synchronous services from smaller, nimbler, and more efficient technologies in the medium to long term.

Australia's NEM experienced "gold-plating" of its network over the first decade of this millennium with overinvestment in capital infrastructure on network providers' regulated asset bases. There is caution in enacting regulation to revisit this through the over-procurement of system strength and synchronous services. The challenge will be in allocating risk and cost appropriately while allowing operator confidence and flexibility within the system to adapt without causing inefficient over-procurement.

A parallel option includes a "Unit Commitment for Security" or "Synchronous Services Market" mechanism that allows the operator to schedule synchronous units to minimum levels for secure operation. The mechanism could then support additional VRE penetration through competitive provision from uncontracted resources. This option can also be potentially supported with a "nomogram" (a diagram that allows calculation through geometrical construction), a precursor of which is the example Transfer Limit Advice table for strength in South Australia (Figure 6: An excerpt of AEMO's Transfer Limit Advice for South Australia, 2020, indicating the combinations of synchronous units [green squares] at low (LOW) levels of system strength that may support various levels of non-synchronous (renewable) generation [column 2]. ). While not exhaustive, this example indicates the various combination of synchronous (gas) units that support different levels of non-synchronous (renewable) generation.

The computational complexity of modeling to construct a table such as this is significant. Additional complexity arises in the inclusion of economic considerations to support efficient decisions in allowing or curtailing renewable energy. Where this economic analysis can be combined with such a table it may provide a pathway forward for a complete nomogram to support greater economic integration of renewables in the short to medium term.

| Combination | Non-sync generation level | Torrens Island A | | | | Torrens Island B | | | | Pelican Point | | | Osborne | | Quarantine or Dry Creek* |
|---|---|---|---|---|---|---|---|---|---|---|---|---|---|---|---|
| | | Ax | Ax | Ax | Ax | Bx | Bx | Bx | Bx | GTx | GTx | ST18 | GT | ST | |
| LOW_2 | ≤ 1,300 MW | | | | | ■ | ■ | | | ■ | | ■ | | | |
| LOW_3 | ≤ 1,700 MW | | | | | ■ | ■ | | | | | | ■ | ■ | ■ |
| LOW_4 | ≤ 1,450 MW | | | | | | | | | ■ | | | ■ | ■ | ■ |
| LOW_5B | ≤ 1,700 MW | ■ | ■ | | | ■ | ■ | | | | | | | | |
| LOW_6 | ≤ 1,700 MW | | | | | ■ | | | | ■ | ■ | | | | |
| LOW_7 | ≤ 1,700 MW | ■ | | | | | | | | ■ | | | ■ | | ■ |
| LOW_8 | ≤ 1,600 MW | ■ | | | | ■ | | | | ■ | ■ | | ■ | | |
| LOW_10 | ≤ 1,750 MW | ■ | ■ | | | | | | | | | | ■ | | |
| LOW_11 | ≤ 1,700 MW | ■ | | | | | | | | ■ | ■ | | | | |
| LOW_13 | ≤ 1,700 MW | ■ | | | | ■ | ■ | | | | | | ■ | | |
| LOW_14 | ≤ 1,300 MW | ■ | ■ | | | ■ | | | | ■ | | ■ | | | |
| LOW_15 | ≤ 1,300 MW | ■ | | | | | | | | ■ | | ■ | | | ■ |
| LOW_18B | ≤ 1,700 MW | ■ | | | | ■ | ■ | | | | | | | | ■ |

*Figure 6: An excerpt of AEMO's Transfer Limit Advice for South Australia, 2020, indicating the combinations of synchronous units [green squares] at low (LOW) levels of system strength that may support various levels of non-synchronous (renewable) generation [column 2]. Ax, Bx etc represent the different generating units of the power stations.*

It is not yet clear how these various approaches may be married nor is it clear how to manage risks and costs of over- and under-procurement to customers via network service providers and/or the system operator. The emerging capabilities of grid-forming inverters (see Box: Australia's Big Battery) will likely play a part in any future mechanism, requiring review and revision over the transition.

## Box: Australia's Big Battery

Following an eight-hour statewide system black event in South Australia in 2016, there was an intense period of government effort to ensure ongoing security for the approaching summer.

Following a series of tweets between billionaires Elon Musk, CEO of Tesla, and Australian Mike Cannon-Brooks, CEO of Atlassian, Tesla offered to build a 100 MW battery within 100 days of contract signing or "it would be free." The South Australian government accepted the offer, subsidizing the initial development cost, expediting planning approvals, and negotiating an ongoing contract for the government to use the battery as an emergency reserve with French developer Neoen as owners of the battery.

In 2017, the Hornsdale Power Reserve (HPR) was commissioned and connected to the grid, becoming the world's largest grid-scale battery at 100 MW/129 MWh. The battery has been a resounding commercial success for both South Australia (SA) customers and Neoen, delivering an estimated $150 million in electricity cost savings to SA customers in its first two, $116 million alone from frequency control costs in a two-week period in 2019 when SA was islanded from the rest of the grid.

The facility has demonstrated the potential of future ESS provision by inverter-connected equipment. The precision with which batteries follow automatic generator control setpoints while providing frequency control ancillary services as compared to a traditional thermal generator is striking (see Figure 10). At present, there is no extra remuneration for facilities that exceed the market ancillary service specification in the NEM. The performance of the battery (typically sub-second) has provided an impetus for the consideration of a fast-frequency response service, which is critical for maintaining security in the power system as inertia levels continue to decrease.

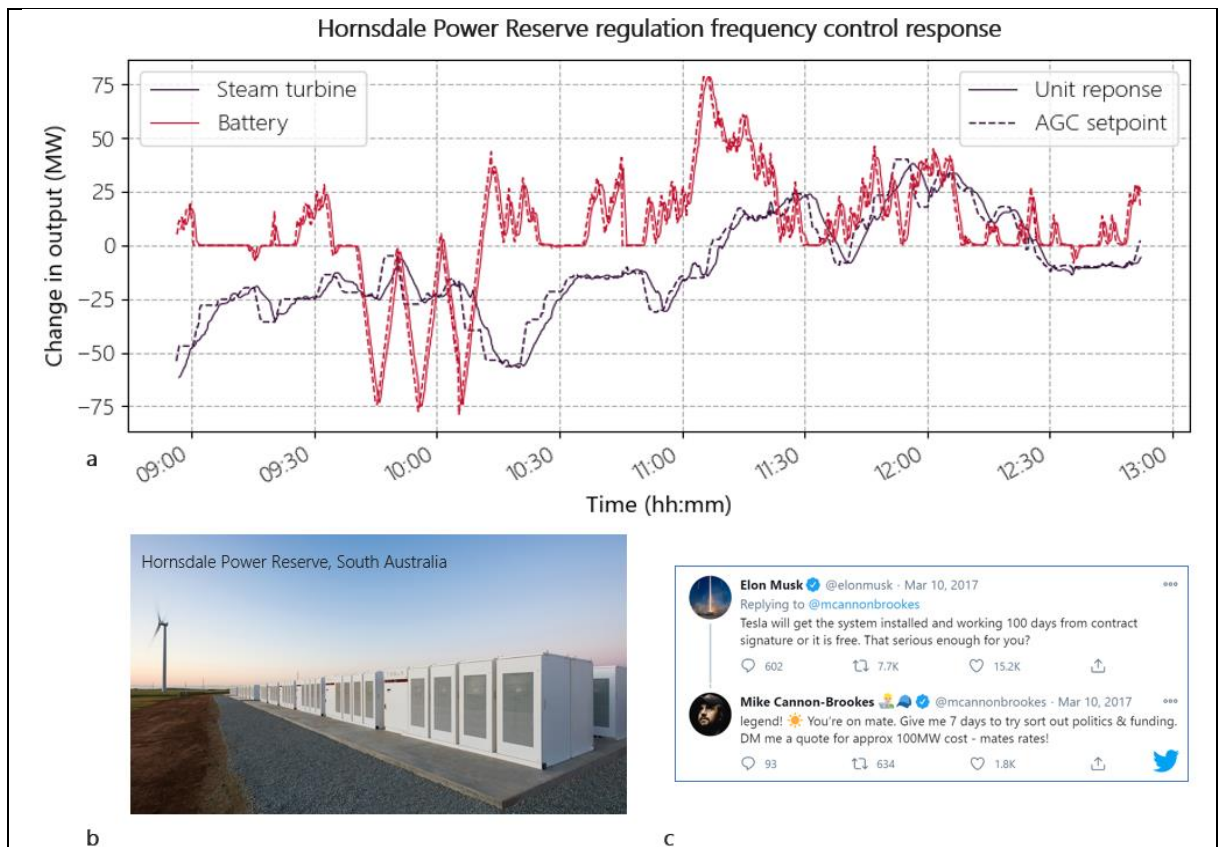

*Figure: a) A comparison of the regulation frequency response capabilities of the Hornsdale Power Reserve compared to a steam turbine;. b) the Hornsdale Power Reserve; c) an extract of the Twitter conversation between Mike Cannon-Brookes and Elon Musk that formed the genesis of the battery.*

In December 2019, the HPR was expanded by 50 MW / 64.5 MWh (total 150 MW/193.5 MWh) with grant and financial support from multiple state and federal initiatives. The upgrade is being delivered by Neoen in collaboration with Tesla, AEMO, and the network service provider ElectraNet to demonstrate the capability of inverter-connected generation to deliver a service equivalent to a synchronously connected generator., which is typically achieved by modeling and implementing the theoretical response of a synchronously connected generator at high speed to govern the response of the facility to power system conditions. Tesla expects to demonstrate an equivalent to 3,000 MWs of functional inertial capability to the system. This capability has not yet demonstrated at grid-scale but may represent a pathway to displacing synchronous generation for provision of these services in the future.

The HPR enjoyed first-mover advantages in becoming the first grid-scale connected battery in Australia. At the time, it was expected to prevent support for additional (N[th] of a kind) battery investment – having assumed to have already taken the majority of available revenue. This has not been the case. At the time of writing, 209 MW of grid-scale battery storage is currently operating. A further 900 MW is expected for delivery by 2024, and 7 GW is in the proposal phase, in addition to several GW of pumped-hydro investments slated across the country.

## Interdependencies

Thermal power stations (largely coal-fuelled) are forecast to retire at pace in the next two decades from both the NEM and the WEM (Figure 7). A key pillar of reform is the consideration of resource adequacy mechanisms to drive investment in capacity to ensure the reliability standard is met through the transition. However, the power system will also need investment in resources capable of meeting the power system's ESS. If these requirements are not considered when investing in new generation (or demand-side) capacity, the overall cost of delivering secure and reliable energy to consumers is likely to be higher.

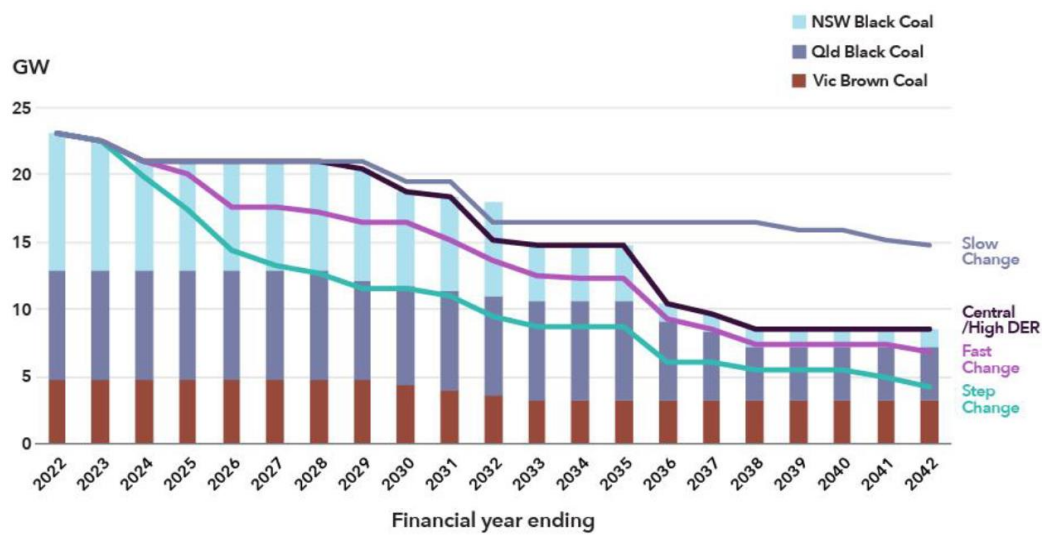

Figure 7: Forecast of coal generation retirements within three Australian states. Source: AEMO ISP 2020

Investment in system service capability may take the form of incremental capital expenditure to new entrant generation, retrofitting existing generators, or new stand-alone merchant resources with system service capabilities. A resource adequacy mechanism (e.g., a capacity market) could be extended to incorporate investment in system service capabilities by placing an obligation on consumers (or retailers) to procure additional capabilities.

These interdependencies present a significant challenge to the overall coordination of reform and market participation over investment horizons. Historically, grid-scale power systems have required large investments in equipment to be economic. In smaller systems and jurisdictions, this has meant that a single provider may be the most economically efficient approach for providing a certain service. Even with rapid DER emergence, it may still hold that a single regulated ESS provider is a more economical solution to open-market provision.

To facilitate investment, the markets for procuring services need to be stable and have clear participation requirements. In theory, markets with sufficient competitive tension will drive efficient investment and retirement decisions, ensuring suitable quantities of each ESS are available. For power systems that lack competitive tension due to either small size or market concentration, a non-market procurement mechanism may be more appropriate. In either case, without appropriately defined services and compensation mechanisms, gaps are likely to appear in the market due to insufficient new investment. Such gaps will not be filled without government or other external intervention.

In particular, DER and demand-side management can likely provide ESS on a cost-competitive basis with traditional and new grid-scale resources. DER can be scaled in a more granular fashion than grid-scale resources once the appropriate rules and initial participation infrastructure are established. This may make them an effective option for augmenting the availability of ESS in multiple time-horizons. As such, it is vital that when revising market arrangements, DER is designed to be part of the solution. If it is not explicitly designed for, there is a real risk that DER won't be able to participate.

The approach needs to balance the requirements of visibility for system operation, distribution-level operation requirements, and implementation cost of any control and communications systems required to facilitate market access. Explicitly considering how DER participates in ESS will allow proponents to build a clear business case and "value-stack" alongside other services to bring the required systems and solutions to market. Without such, mechanisms could drive separate capital investments to meet each of the power system requirements, increasing costs to consumers.

## The Australian Approach and the Future of Essential System Services

Australia's electricity system is rapidly transitioning from a generation fleet dominated by coal and gas to accommodating the world's highest penetration of residential solar PV (22% of all stand-alone houses) with regular instantaneous provision of 100% renewable supply likely within five years. This will occur on the east coast with a grid covering more than three times the area of Texas and in southwestern Australia across an area the size of the United Kingdom.

Catalyzed by the rapid pace of change and through a handful of significant system security events, Australian governments have instigated sweeping market reforms to support the transition to higher VRE penetration. A key focus is on ESS with recognition of services once provided by synchronous generators as a byproduct of energy generation and not yet replaced by inverter-based technologies.

Although there are regulatory and physical differences between the west and east coast markets, the philosophical and economic principles established during their conception have been maintained. Included are the importance of efficient price signals in operational timeframes based on voluntary bids and offers, facilitating overall dispatch while maximizing market-based outcomes and minimizing interventions.

Regarding the reform of specific system services, Figure 8: A possible roadmap for ESS in Australia (NEM and WEM) to 2025 and beyond, indicating an evolution towards spot market based mechanisms where possible. Adapted from FTI Consulting's Report to the ESB 2020. outlines a graphical roadmap indicating the pathways for reform in both markets.

For the NEM, this involves the possible implementation of:

- A new operating reserve spot market likely based on a five-minute or 30-minute ramping availability product procuring either the total ramp or holding reserve out of the market with a separate call mechanism.
- A new fast frequency response market (sub two seconds) to encourage and reward provision of rapid frequency control from batteries, and refinement of the mandatory requirement for primary frequency control (recently enacted and already delivering market improvements to system-wide frequency performance).
- A new framework for system strength where the system operator sets minimum/efficient levels of strength (via a short-circuit current ratio) at all nodes of the network, and the network service provider is obliged to maintain these levels. There will likely be a mechanism to schedule synchronous resources in operational timeframes to provide inertia and system strength with support for longer-term consideration of an inertia spot market.

For the WEM, the reform pathway includes:

- A new spot market for regulation frequency management, and transformation of the current contingency frequency control framework to spot markets similar to the NEM.
- The implementation of a ROCOF control service spot market to pay for inertia in operational timeframes, the first such market we are aware of anywhere in the world.

For all new services, there is an explicit awareness of the importance of setting technical requirements to support and encourage emerging technologies and, in particular, possible future DER capabilities and demand-side participation.

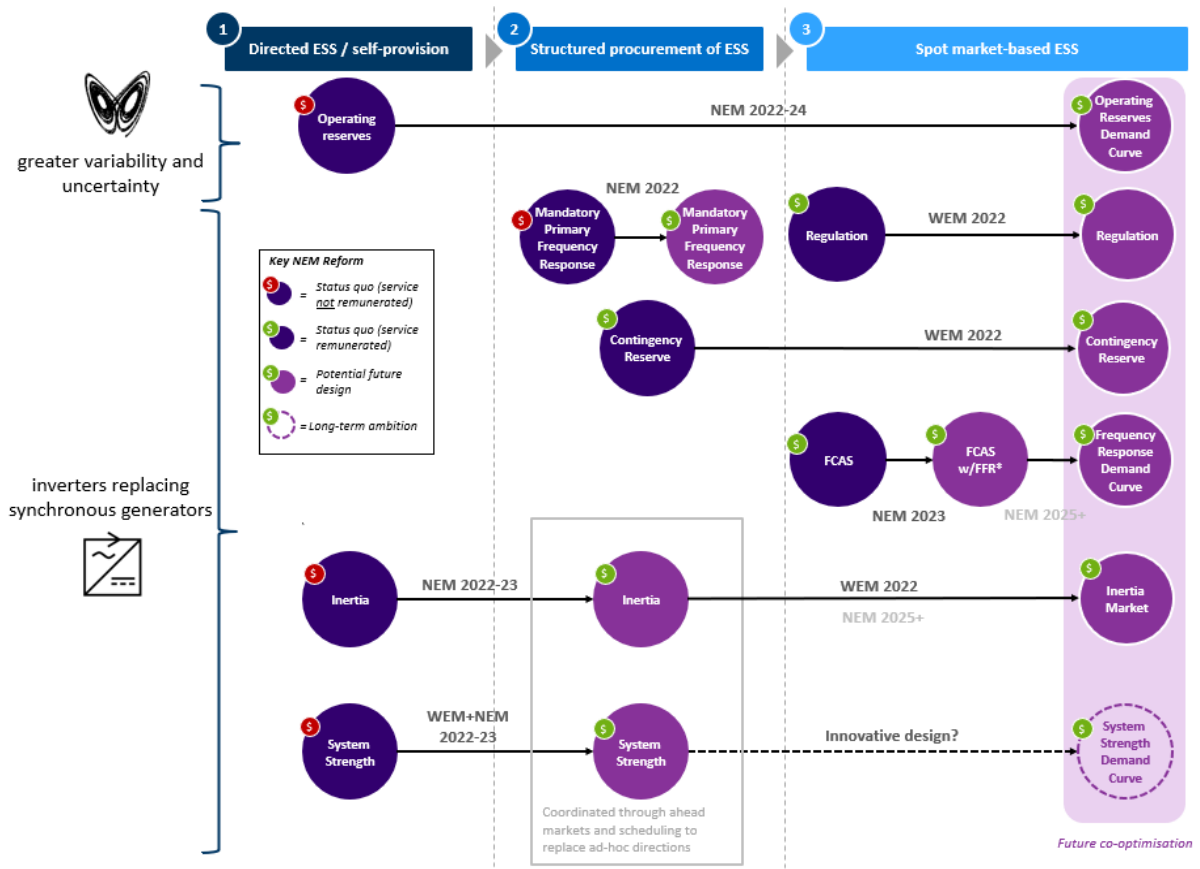

*Figure 8: A possible roadmap for ESS in Australia (NEM and WEM) to 2025 and beyond, indicating an evolution towards spot market based mechanisms where possible. Adapted from FTI Consulting's Report to the ESB 2020.*

Both NEM and WEM reform programs are ongoing. Development to date has required robust collaboration across market operators, regulators, and government agencies, and extensive engagement with market participants, including generators, retailers, DER aggregators, consumer representatives, and network operators.

The rapid pace of change has been catalysed by legislated net-zero emission targets from states and territories towards 2050, though a national target has not yet been set. Australia is the world's largest exporter of both coal and natural gas, ensuring the impact of measures to address climate change on these industries generate significant political debate with extensive business lobbying. This may, in part, explain why Australia has struggled over the past two decades to navigate a middle path of the electricity transition with bipartisan support.

But even with uncertain support at a federal level, and perhaps in part because of it, Australian households have embraced rooftop solar at world-leading levels, and industrial buildings are now following. Spurred by broad political support at the state level for net-zero targets, state and territory governments are investing heavily in renewable generation through reverse auctions, and power purchase agreements. They are making investments in transmission design flagged by the system operator as essential to support emerging "'renewable energy zones." These zones are discussed in another article, "Planning at System Level, Renewable Energy Zones."

Reform programs are underway, but with significant work still to be completed. For both the NEM and WEM, the detailed work of market design, technical qualification, compliance, and regulatory frameworks has yet to be finalized. Each will have a significant impact on market participant behaviour and system outcomes, and there is a growing recognition of the value in allowing flexibility to those involved in the transition. Australia is likely to continue on its reform pathway for the coming decade due to the rapid pace of change in both supply and demand.

The current reform of ESS predominantly addresses challenges arising from the inverter-based replacement of synchronous generation with early steps to addressing the emerging variability and uncertainty of supply. Future essential services will likely be needed to a) mitigate minimum demand (already a pressing security concern for some regions; b) provide individual components of system strength (where fungible; and c) provide broader provision of system restart services to support greater resilience and "islandability" in the event of bushfires and extreme weather events.

All future reforms will need to interact fairly with DER, recognising that the advanced grid-forming technological capabilities of new battery technologies, such as the Hornsdale Power Reserve, will likely be eventually translated to the power electronics of smaller inverters at the household scale. To support customer participation and fairness, this may be facilitated through a broadly agreed "DER Bill of Rights" with principles that could include 1) the allowance of near-unimpeded self-consumption of self-generated electricity (even if export may be curtailed); 2) the fair imposition of technical requirements to support grid security; and 3) remuneration for energy and system services proportional to that received by large-scale resources.

As the electrification of transport proceeds at pace alongside increased sophistication of demand-side participation, there will likely be new system service needs and opportunities for provision from emerging resources such as electric cars. This will need to be accompanied by a redefinition of new roles for network service providers.

As the energy transition gathers momentum through the new millennium, Australia finds itself rapidly departing from the paradigm first enacted in 1899 of default system service provision from synchronous resources. It is moving towards new market frameworks that remunerate the provision of distinct services in real-time from technology unimaginable 100 years ago. How Australia addresses this change has the potential to help inform the global energy transition in the coming century for the urgent decarbonization journeys which all countries across the world are now navigating.

## For Further Reading

- Blakers, Stocks and Baldwin, 2020 "Australia, the global renewable energy pathfinder." ANU Energy Institute
- AEMO Renewable Integration Study Stage 1 Report, 2020
- https://aemo.com.au/en/energy-systems/major-publications/renewable-integration-study-ris
- Finkel. A, 2017, Independent Review into the Future Security of the National Electricity Market - Blueprint for the Future, Office of the Chief
- The Australian Energy Security Board Post 2025 Market Electricity Design Project https://esb-post2025-market-design.aemc.gov.au/

- AEMO Integrated System Plan, 2020 https://aemo.com.au/en/energy-systems/major-publications/integrated-system-plan-isp/2020-integrated-system-plan-isp

## Biographies

- Niraj Lal is with the Australian Energy Market Operator and Australian National University, Australia.
- Toby Price is with the Australian Energy Market Operator, Australia.
- Leon Kwek is with the Australian Energy Market Operator, Australia.
- Christopher Wilson is with the Australian Energy Market Operator, Australia.
- Farhad Billimoria is with the Australian Energy Market Operator, Australia, and Oxford University, United Kingdom.
- Trent Morrow is with the Australian Energy Market Operator, Australia.
- Matt Garbutt is with the Energy Security Board, Australia.
- Dean Sharafi is with the Australian Energy Market Operator, Australia.